\documentclass[prl,twocolumn,superscriptaddress,showpacs,debugon,nofootinbib]{revtex4}
\usepackage{epsfig}
\newcommand{\bk}{{\bf k}}
\newcommand{\be}{\begin{equation}}
\newcommand{\ee}{\end{equation}}
\renewcommand{\r}{{\bf r}}
\newcommand{\s}{{\bf s}}

\newcommand{\q}{{\bf q}}

\newcommand{\R}{{\bf R}}

\begin{document}


\title{Intensity correlations and mesoscopic fluctuations of diffusing photons in cold atoms}
\author{O. Assaf and E. Akkermans}
\affiliation{Department of Physics, Technion Israel Institute of Technology,
  32000 Haifa, Israel}

\begin{abstract}

We study the angular correlation function of speckle patterns that result from multiple scattering of photons by cold atomic clouds. We show that this correlation function becomes larger than the  value given by Rayleigh law for classical scatterers. These large intensity fluctuations constitute a new mesoscopic interference effect specific to atom-photon interactions, that could not be observed in other systems such as weakly disordered metals. We provide  a complete  description of this behavior and  expressions that allow for a quantitative comparison with experiments.
\end{abstract}


\date{\today}

\maketitle


A wave propagating in a random medium undergoes multiple
scattering and the intensity pattern resulting
from  interferences of the scattered waves with each other is
known as a speckle pattern. The angular and time dependent
properties of these patterns have been extensively studied
\cite{shapiro,am, maret}.  They exhibit coherent mesoscopic
effects and provide a sensitive probe to  scattering properties of
diffusive media. Quasi-resonant elastic scattering of photons by
cold atomic gases represents in this context an important issue,
since it provides a new tool to study properties of cold atomic
gases such as atomic dynamics. Photon  propagation in atomic gases
differs  from the case of electrons in disordered metals
\cite{rama} or of electromagnetic waves in suspensions of
classical scatterers, due to the existence of atomic internal
degrees of freedom coupled  to the photon polarization. Some
effects  of a  Zeeman degeneracy on the coherent backscattering
\cite{am,akkmayn} have  been recently investigated  in the weak
scattering  limit \cite{miniat} in terms of phase coherence times
\cite{amm}.

The purpose of this work is to study the {\it static angular
correlation function} of photons performing coherent multiple
scattering in a cold atomic gas. The photon intensity correlation
function between angular scattering channels is defined using the
transmission coefficient ${\cal{T}}_{ab}$ by
 \be C_{aba'b'} = {\overline{\delta {\cal T}_{ab} \delta {\cal T}_{a'b'}} \over \overline {\cal T}_{ab} \overline {\cal T}_{a'b'}} \, \ .
\label{1217} \ee Here, $\overline{\cdot \cdot \cdot}$ denotes a
configuration average over both the position of atoms and their internal
degrees of freedom (see below) and $\delta {\cal T}_{ab}  \equiv  {\cal T}_{ab} - {\overline
{\cal T}}_{ab}$. For classical scatterers,
intensity fluctuations obey the Rayleigh law $C_{abab} =1$. In the
presence of a Zeeman degeneracy, angular correlations of speckle
patterns and intensity fluctuations become larger than one.  This
is a new and genuine mesoscopic effect specific to multiple
scattering of photons by atoms and directly related to interference between amplitudes associated to different atomic quantum states.

Atoms are modeled as degenerate two-level systems denoted by
$|j_{g} m_{g} \rangle$ for the ground state and $|j_{e} m_{e}
\rangle$ for the excited state, where $j$ is the total angular
momentum and $m$ is its  projection on a quantization axis. The levels
are degenerate with $|m_g| \leq j_g$ and $|m_e| \leq j_e$.

We refer to the following possible experimental setup (Fig.1). A
light pulse is incident along a direction ${\hat \s}_a$ onto a dense
enough atomic gas confined in a slab geometry. This pulse is detected
along a direction ${\hat \s}_b$ after being multiply scattered
($ab$ channel). A time $\tau$ later, a second pulse that
corresponds to the $a'b'$ channel, is detected. We assume that the
time $\tau$ is short enough so that the atoms stay at rest
between the two pulses. The same measurement is repeated
after a time $T \gg \tau$, during which the scatterers move. The
averaging over spatial disorder results from this motion. This is
a Young-like experiment. Thus, although a pulse contains {\it
many} photons, the transmitted intensity ${\cal T}_{ab}$  is
proportional to the probability of {\it one} ``representative"
photon incoming along ${\hat \s}_a$, to emerge along ${\hat
\s}_b$.

\begin{figure}[ht]
\centerline{ \epsfxsize 9cm \epsffile{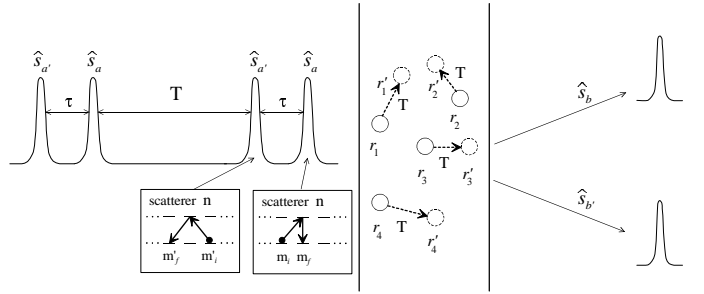} } \caption{\em
Photons in each {\it pair} of pulses are
scattered by atoms at identical positions $\r_i$ but with
distinct and uncorrelated quantum numbers $(m_i , m_f )$ and $(m_i
' , m_f ' )$.  After a time $T \gg \tau$, a new measurement is
performed after the atoms moved to new positions $\r_i '$. }
\label{FigureOhad}
\end{figure}

The average transmission coefficient $\overline{\cal T}_{ab}$ is
obtained by summing all the possible scattering amplitudes,
$A_n^{\{R,m\}}$, corresponding to a given configuration $\{R,m\}$.
Here $\{R\}$ accounts for the spatial positions of all scatterers,
and $\{m\}$ is a notation for their internal Zeeman states both
before and after scattering. The index $n$ denotes {\it one}
possible multiple scattering path. Squaring the sum of amplitudes
we have \cite{rk2} \be \overline{\mathcal{T}}_{ab}= \overline{
|\sum_{n}A_n^{\{R,m\}}|^2} = \sum_{n
n'}\overline{A_n^{\{R,m\}}A_{n'}^{\{R,m\}*}}  \ee where
$\overline{\cdot \cdot \cdot}$ denotes a configuration average, over both $\{R\}$ and $\{m\}$. When averaging over $\{R\}$,
all cross terms $n\neq n'$ vanish because of large fluctuating
phase shifts, so that $\overline{\mathcal{T}}_{ab}=\sum_{n}
\overline{ |A_n^{\{m\}}|^2}$. This expression, known as the {\it intensity Diffuson}, is the leading approximation in the weak
disorder limit $k_0 l \gg 1$, where $k_0$ and $l$ are respectively
the wave number and the  elastic mean free path of photons. The
$(ab)$ pulse contains many photons, and each of them may change the
internal state of atoms. Therefore, if $\{m\}$ (resp. $\{m'\}$) is
the atomic internal configuration seen by a ``representative"
photon of the $ab$ (resp. $a'b'$) pulse, then we can assume that
there is no correlation between $\{m\}$ and $\{m'\}$. 

Similarly to
the average intensity (2), the correlation of the transmission
coefficient is
\begin{eqnarray}
\overline{{\cal T}_{ab} {\cal T}_{a'b'}} &=& \overline{\mathcal{T}_{ab}^{\{R,m\}}
\mathcal{T}_{a'b'}^{\{R,m'\}}}  \nonumber \\
&=& \overline{\sum_{ijkl}A_i^{\{R,m\}}A_j^{\{R,m\}*}A_k^{\{R,m'\}}A_l^{\{R,m'\}*}}
\end{eqnarray}
since,  as before, the averaging over $\{R\}$ leaves only pairs of
amplitudes having exactly opposite phase shifts. To leading order
in the weak disorder limit, the only non vanishing contributions
involve two {\it Diffusons}, {\it i.e.} two possible pairings of
amplitudes, either $i=j,k=l$, which gives
$\overline{\mathcal{T}}_{ab} \overline{\mathcal{T}}_{a'b'}$, or
$i=l,j=k$ so that
 \be
\overline{\delta \mathcal{T}_{ab} \delta \mathcal{T}_{a'b'}} =
\sum_{i j} \overline{
A_i^{\{m\}}A_i^{\{m'\}*} A_j^{\{m'\}}A_j^{\{m\}*}}  \, \ .
\ee The correlation function thus appears as products of two amplitudes, that correspond to different internal configurations $\{m\}$ and $\{m'\}$, but to
the same scattering path $i$ (or $j$). Most of multiple scattering paths $i$ and $j$ do not share common  scatterers so that  we can average $A_i^{\{m\}}A_i^{\{m'\}*}$ and
$A_j^{\{m'\}}A_j^{\{m\}*}$ separately, since these averages are
taken upon different atoms, and finally,
\be
\overline{\delta \mathcal{T}_{ab} \delta \mathcal{T}_{a'b'}} =  \Bigl| \sum_i \overline{
A_i^{\{m\}}A_i^{\{m'\}*}} \Bigr|^{2}
\label{correl1}
\ee

In the theory of multiple scattering it is helpful to use a
continuous description \cite{am}. In this
framework, one defines two {\it Diffuson}
functions $\mathcal{D}^{(i,c)}$ by \cite{rk2}
\be
\overline{\cal T}_{ab} = \int d \r d \r'
 {\cal D}^{(i)} (\r, \r') \,
\label{intensitedalbedo12}
\ee
and
\be
\overline{\delta {\cal T}_{ab} \delta {\cal T}_{a'b'}} = \Bigl|
\int d \r d \r' e^{ik_0 [\Delta {\hat \s}_{a}.\r - \Delta {\hat
\s}_{b}.\r']}  {\cal D}^{(c)} (\r, \r') \Bigr|^2 \label{corramplitude}
 \ee
where $\Delta {\hat \s}_{a,b} = {\hat \s}_{a,b} - {\hat
\s}_{a',b'}$. The {\it intensity Diffuson}
$\mathcal{D}^{(i)}(\mathbf{r},\mathbf{r}')$ is the sum of the
terms $\overline{
|A_n^{\{m\}}(\mathbf{r},\mathbf{r}')|^2} $ between
endpoints $\mathbf{r}$ and $\mathbf{r}'$.  On the other
hand, the {\it correlation Diffuson}
$\mathcal{D}^{(c)}(\mathbf{r},\mathbf{r}')$ is the sum of the terms $\overline{A_i^{\{m\}}(\mathbf{r},\mathbf{r}')
A_i^{\{m'\}*}(\mathbf{r},\mathbf{r}')} $, {\it
i.e.}, that involve uncorrelated configurations $\{m\}$
and $\{m'\}$.

The two functions ${\cal D}^{(i,c)}$ are obtained from the
iteration of a proper elementary vertex ${\cal V}^{(i,c)}$, that
describes the microscopic details of the scattering process. The
iteration of the elementary vertex is written symbolically (either
for ${\cal D}^{(i,c)}$,${\cal V}^{(i,c)}$ we shall denote by
${\cal D}$,${\cal V}$) as \be {\cal D} =  {\cal V} + {\cal V}
{\cal W} {\cal V} + \cdots  = {\cal V} + {\cal D} {\cal W}  {\cal
V} \, \ .\label{iterationD} \ee The term ${\cal V} $ accounts for a
single scattering and $ {\cal D} {\cal W} {\cal V}$ represents its
iteration. The quantity $\cal W$ describes the propagation of the
photon intensity between successive scattering events and it will
be described later on.

The elementary vertex is obtained by the pairing of two scattering
amplitudes of a photon by an atom. It is given by \cite{rose}
\be
 {\cal V} = \frac{4 \pi/ l}{2j_g+1} \sum_{m_i}\langle
j_g m_2|V(\hat {\bf \varepsilon}_1,\hat {\bf \varepsilon}_2)|j_g
m_1\rangle  \langle j_g m_4|V(\hat {\bf \varepsilon}_3,\hat {\bf
\varepsilon}_4)|j_g m_3\rangle^\ast \label{vertexeq} \ee where the operator $ V(\hat {\bf
\varepsilon}',\hat {\bf \varepsilon}) = \sum_{m_e} \hat {\bf
\varepsilon}'^{*} \cdot{\bf d}|j_e m_e \rangle\langle j_e m_e | {\bf
d}  \cdot \hat {\bf \varepsilon} $ results from the dipolar
interaction energy $ -{\bf d}.{\bf E}$ between atoms and photons,
$\bf d$ being the atomic dipole operator and $\bf E$ the electric
field operator. In the case of ${\cal V}^{(c)}$, each one of the two
coupled scattering amplitudes in (\ref{vertexeq}) might belong to a
distinct atomic configuration, meaning that we must consider two
distinct couples of initial $(|j_g m_1\rangle , |j_g m_3 \rangle)$
and final $(|j_g m_2\rangle , |j_g m_4 \rangle)$ atomic states, as
well as two initial $(\hat {\bf \varepsilon}_1, \hat {\bf
\varepsilon}_3)$ and final $(\hat {\bf \varepsilon}_2, \hat {\bf
\varepsilon}_4)$ photon polarization states. The summations
over the quantum numbers $m_i$ result from averaging over initial
atomic states and from non detected final states. Thus ${\cal
V}^{(c)}$ corresponds to the most general case regarding the $m_i$
quantum numbers. In contrast, ${\cal V}^{(i)}$ corresponds to the differential scattering cross-section, for which we set $m_1 =
m_3$, $m_2 = m_4$, $\hat {\bf \varepsilon}_1 = \hat {\bf
\varepsilon}_3$ and $\hat {\bf \varepsilon}_2 = \hat {\bf
\varepsilon}_4$ in (\ref{vertexeq}). This is because the intensity Diffuson is built out of two coupled amplitudes that must belong to the same
scattering process. This distinction between ${\cal V}^{(i)}$ and
${\cal V}^{(c)}$, and therefore between ${\cal D}^{(i)}$ and ${\cal
D}^{(c)}$, occurs only for $j_g > 0$ and it is at the basis of the
new results we obtain here for mesoscopic speckle
correlations.

The iteration (\ref{iterationD}) is implemented using the
decomposition of the various terms into {\it standard basis}
components, thus leading to the definition of rank four tensors
such as \be {\cal V} = \sum_{\alpha \beta \gamma \delta} (\hat
{\bf \varepsilon}_1)_{-\alpha} (\hat {\bf \varepsilon}_2)_\gamma
^* (\hat {\bf \varepsilon}_3)_{-\beta} ^* (\hat {\bf
\varepsilon}_4)_\delta \, \ {\cal V}_{\alpha \beta, \gamma \delta}
\label{vcomposantes} \ee Likewise, the iteration equation
(\ref{iterationD}) for the Diffusons acquires a tensorial
structure, which reads
 \be
 {\cal D}_{\alpha \beta, \gamma \delta} =  {\cal V}_{\alpha \beta, \gamma \delta}  + W \sum_{\mu \nu \rho \sigma}  {\cal D}_{\alpha \beta, \mu \nu} b_{\mu \nu , \rho \sigma} {\cal V}_{ \rho \sigma , \gamma \delta} \, \ .
\label{it2} \ee Here ${\cal W}=W b$, the function $W$ describes the scalar part of the photon intensity propagator and $b$, defined by \be b_{\alpha \beta, \gamma
\delta} = \langle \left(\delta_{\alpha\gamma} - (-)^{\gamma}{\hat
s}_\alpha {\hat s}_{-\gamma} \right) \left(\delta_{\beta\delta} -
(-)^{\beta}{\hat s}_{-\beta} {\hat s}_\delta \right) \rangle ,
\label{tensorb} \ee accounts for the polarization dependent part.
This expression follows at once by noticing that after being
scattered by an atom, the two outgoing photon amplitudes propagate
with a wavevector ${\hat {\s}} = \bk / k_0$, random in direction
but identical for both, and with two different polarization
components. Since ${\hat {\s}}$ is random, the intensity
propagation is averaged $\langle \cdot \cdot \cdot \rangle$ over
photon wavevectors direction. The term $\delta_{\mu\nu} -
(-)^{\nu}{\hat s}_\mu {\hat s}_{-\nu}$ expresses transversality.

To proceed further, we use the Wigner-Eckart theorem to rewrite
the tensor ${\cal V}_{\alpha \beta, \gamma \delta}$ in terms of a
summation of product of $3j$-symbols,
\begin{eqnarray}
&&{\cal V}_{\alpha \beta, \gamma \delta} = 3 (2j_e +1) a_{j_g j_e}
\sum_{m_i m_e m_e '} \left(
\begin{array}{clcr}j_e &
1& j_g \\
-m_e & \alpha & m_1 \end{array} \right) \times \nonumber \\
&\times& \left(
\begin{array}{clcr}j_e &
1& j_g \\
-m_e & \gamma & m_2 \end{array} \right)  \left(
\begin{array}{clcr}j_e &
1& j_g \\
-m_e ' & \delta & m_4 \end{array} \right) \left(
\begin{array}{clcr}j_e &
1& j_g \\
-m_e ' & \beta & m_3 \end{array} \right)
\label{vertexgeneraliseabcd}
\end{eqnarray}
where $a_{j_g j_e} = (2 j_e +1) / 3 (2 j_g +1)$. The two tensors
$b_{\alpha \beta, \gamma \delta}$ and  ${\cal V}_{\alpha \beta,
\gamma \delta}$ can be written in the form of a $9 \times 9$
matrix. According to the spectral decomposition theorem, they can
be decomposed using an orthonormal set of (generally) nine
projectors $T^{(K)}$ \cite{ohaderic}.
Looking at (\ref{it2}), we wish to find the spectral decomposition
of $\cal D$ using the spectral decomposition of $\cal V$ and $b
{\cal V}$. The problem is that they do not share the same
projectors set in their spectral decomposition. We are thus led to
define a new tensor $U$ by $ {\cal D} = U {\cal V}$. It obeys the
iteration equation $U = 1 + W U {\cal V} b$ and it involves only
the spectral decomposition of ${\cal V}^{(i,c)} b=\sum_{K=0}^8
u^{(i,c)}_K T^{(K)}$. This leads immediately to
 \be
 {\cal D}_{\alpha \beta, \gamma \delta} ^{(i,c)} = \sum_K U_K ^{(i,c)} \left({\cal V}_K ^{(i,c)} \right)_{\alpha \beta, \gamma \delta}
\label{eq13} \ee with $ {\cal V}_K ^{(i,c)} = T^{(K)} {\cal
V}^{(i,c)}$ and \be U_K ^{(i,c)} = {4 \pi / l \over 1 - W(q) u_K
^{(i,c)}} \simeq  { 8 \pi c \over 3 l^2}  a_{j_g j_e} {1 /  u_K
^{(i,c)} \over {1 \over \tau_K ^{(i,c)}} + D q^2 } \label{eq14}
\ee where $\q$ (with $q=|\q|$) is the Fourier variable of the
difference $\R=\r'-\r$ between the two endpoints of a multiple
scattering sequence. The {\it r.h.s} in (\ref{eq14}) is obtained
by using the diffusion approximation ({\it i.e.} $ql \ll 1$), so
that $W(q) \simeq {3 \over 2 a_{j_g j_e}} (1 - q^2 l^2 / 3)$,
where $D= c l / 3$ is the photon diffusion coefficient \cite{am} and $c$ the speed of light.
We identify the set of characteristic times \be \tau_K ^{(i,c)} =
\left( {l \over c} \right) { u_K ^{(i,c)} \over {2 \over 3 } a_{j_g j_e} - u_K
^{(i,c)}} \, \ . \label{tauk} \ee For ${\cal V}^{(i)}$ ($m_1 =
m_3$, $m_2 = m_4$ in (\ref{vertexgeneraliseabcd})), it is
straightforward to check that $({\cal V }^{(i)} b)_{\alpha \beta,
\gamma \delta}$ admits a spectral decomposition over 3 projectors
$T^{(K)}$ only \cite{am,miniat,mmam}. In contrast, for $ {\cal
V}^{(c)}$ there are no constraint on the $m_i$ quantum numbers, so
that the total angular momentum needs not to be conserved, and the
corresponding spectral decomposition involves usually more than 3
projectors $T^{(K)}$. For a non degenerate ground state level
($j_g = 0$), angular momentum is automatically conserved and the
two vertices become identical, $ {\cal V}^{(c)} = {\cal V}^{(i)}$.

The poles that occur in (\ref{eq14}) correspond to diffusive modes
of lifetime $\tau_K ^{(i,c)}$. This shows up when rewriting
(\ref{eq13}), with the help of (\ref{vcomposantes}), in real space
\be {\cal D}^{(i,c)} (\r,\r') =   \sum_K {Y_K ^{(i,c)} \over u_K
^{(i,c)}} \int_0^{\infty} dt  {\cal D}(\r,\r',t) e^{-
t/ \tau_K ^{(i,c)}}  \label{it3} \ee where \be Y_K ^{(c)} =
\sum_{\alpha \beta \gamma \delta}  ({\hat {\bf
\varepsilon}}_a)_{-\alpha}({\hat {\bf \varepsilon}}_b)_\gamma ^*
({\hat {\bf \varepsilon}}_{a'})_{-\beta} ^* ({\hat {\bf
\varepsilon}}_{b'})_\delta \left({\cal V}_K ^{(c)} \right)_{\alpha
\beta, \gamma \delta} \ee and a corresponding expression for $Y_K
^{(i)}$ obtained by setting $a'=a$, $b' = b$ and ${\cal V}^{(i)}$
in the previous relation.
The scalar Diffuson propagator ${\cal D}(\r,\r',t)$ obeys a
diffusion equation whose solution for a slab of width $L$ is well known
\cite{am,ohaderic} and leads for (\ref{intensitedalbedo12}) and
(\ref{corramplitude}) to \be {\overline {\cal T}}_{ab} =
\sum_{K=0}^2 {Y_K ^{(i)} \over u_K ^{(i)}} {\cal D} \left(Q_K
^{(i)} (0) \right) \label{avt} \ee and \be  \overline{ \delta {\cal T}_{ab} \delta {\cal T}_{a'b'}}  = \delta_{\Delta {\hat \s}_{a},\Delta {\hat \s}_{b}}
\Big[ \sum_{K=0}^8 {Y_K ^{(c)} \over u_K ^{(c)}} {\cal D} \left(
Q_K ^{(c)} (q_p) \right) \Big]^2  \, . \label{corr} \ee We have defined
the quantities $q_p =k_0 \Delta {\hat \s}_{a}$, $ Q_K ^{(i,c)}
(x) = \sqrt{x^2 + (1 /  D \tau_K ^{(i,c)})} $ and ${\cal D} (x) =
\sinh^2 (x  l) /  (x l  \sinh (x L))$.

\medskip

We now analyze expressions (\ref{avt}) and (\ref{corr}), which
constitute the main results of this paper. First, consider the
modes of the average intensity ${\cal D}^{(i)}$. It is easy to
check that $\tau_0 ^{(i)}$ is infinite as a result of the Ward
identity $u_0 ^{(i)} =  {2 \over 3 } a_{j_g j_e}$ (see
(\ref{tauk})). The corresponding Goldstone mode $U_0 ^{(i)}
\propto 1/ D q^2$ expresses energy conservation and long-range
propagation of the average intensity. The two other modes $U_K
^{(i)}$ are exponentially damped with the times $\tau_K ^{(i)}$
(see (\ref{it3})). This expresses photon depolarization in
multiple scattering \cite{miniat,am,amm,omont}.

The spectral decomposition of ${\cal D}^{(c)}$ gives rise to  nine
modes and their  corresponding times $\tau_K ^{(c)}$. Such times
are well-known to occur in quantum mesoscopic physics {\it e.g.}
in conductance fluctuations of disordered metals in the presence
of magnetic impurities \cite{am,glazman}. The times $\tau_K
^{(c)}$ describe how underlying interferences between multiply
scattered waves (electrons in metals, photons in the present case)
are washed out in the presence of other degrees of freedom. The
surprising and new feature of the atom-photon scattering, is the
occurrence of a  mode $(K=0)$ with a negative $\tau_0 ^{(c)}$.
According to (\ref{it3}), this corresponds to an amplified mode
that enhances the angular correlation function. This amplified
mode is present for a degenerate atomic transition ($j_g,j_e>0$)
and vanishes otherwise. Its origin can be traced out from the
vertex (\ref{vertexeq}) which can be written as a sum of an
incoherent contribution ({\it i.e.}, a sum of probabilities)
present both in ${\cal V}^{(i)}$ and ${\cal V}^{(c)}$, and a
coherent contribution ({\it i.e.}, a sum of products of quantum
amplitudes associated to interferences between different Zeeman
states). The coherent contribution enhances ${\cal V}^{(c)}$ with
respect to ${\cal V}^{(i)}$ and its iteration gives rise to the
amplified mode characterized by $\tau_0 ^{(c)}$.
\begin{figure}[ht]
\centerline{ \epsfxsize 6cm \epsffile{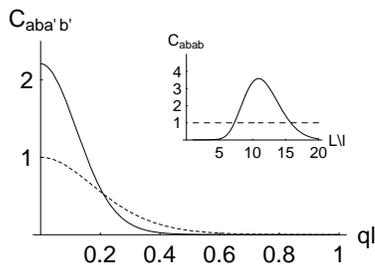} } \caption{\em
Angular correlation function $C_{aba'b'}$ plotted as a function of
$q = k_0  \Delta {\hat \s}_{a}$ for an atomic transition
between non degenerate ($j_g=0$,$j_e=1$, dashed line) and degenerate
energy levels ($j_g=1$,$j_e=2$, solid line). The non degenerate case
describes the clasical Rayleigh scattering of a polarized wave. These curves
correspond to $L = 7 l_e$
. The
inset gives the dependence of intensity fluctuations $C_{abab}$ upon
the width $L$ (the dashed line is the Rayleigh law).} \label{fig4}
\end{figure}
Expressions (\ref{avt}) and (\ref{corr}) lead to an expression of $C_{aba'b'}$
plotted in Fig.\ref{fig4}. This expression allows to recover the
limiting case of a scalar classical wave \cite{shapiro}
corresponding to $Y^{(c)} = Y^{(i)} =1$ and to a single mode with
infinite $\tau_0 ^{(i,c)}$. It also provides a simple expression
for angular speckle correlations of classical Rayleigh scattering
of a polarized wave \cite{ohaderic,genak}. For atomic transitions
between degenerate levels, $C_{aba'b'}$ exhibits a steep decrease
and large intensity fluctuations (see Fig.\ref{fig4}) as compared
to the case of non degenerate atomic levels. Intensity
fluctuations measured by $C_{abab}$ become larger than one,
unlike Rayleigh law, $C_{abab} =1$, well-obeyed by classical
scatterers.
Large intensity fluctuations result from the amplified mode
$\tau_0 ^{(c)}$ which leads to a divergence of the integral in
(\ref{it3}). This divergence is cutoff by other dephasing
mechanisms, such as Doppler shift, inelastic scattering or finite
size of the atomic trap. Denoting by $\Lambda$ this upper cutoff,
and assuming that the dominant contributions to $\overline {\cal
T}_{ab}$ and to $\overline{\delta {\cal T}_{ab} ^2}$ are given
respectively by the Goldstone and the amplified modes, we deduce
from (\ref{it3}) that \be C_{abab} =  {\overline{\delta {\cal
T}_{ab} ^2} \over \overline {\cal T}_{ab} ^2} \simeq  A \pi^4
\left( {e^{X \Lambda D / L^2} - 1 \over X} \right)^2
\label{relfluc} \ee where $X \equiv (L / L_0 ^{(c)})^2 - \pi^2$ involves  the diffusion length $L_0 ^{(c)} \equiv \sqrt{D
|\tau_0 ^{(c)}|}$ and $A$ is a constant equal to one for a non
degenerate ground state. This approximate expression reproduces
the main features of $C_{abab} $ plotted in the inset of
Fig.\ref{fig4}, namely that it can be larger than the Rayleigh
term and that it is peaked at a value of $L$ that depends on the
cutoff $\Lambda$.  When $L_0 ^{(c)}$ and $\Lambda$ are infinite,
$C_{abab}$ becomes independent of $L$ and is given by the Rayleigh law. Relative fluctuations as given by (\ref{relfluc}) thus
provide a direct probe of  dephasing mechanisms in cold atomic
gases.

This research is supported in part by the Israel Academy of
Sciences and by the Fund for Promotion of Research at the
Technion.


\begin{thebibliography}{99}
\bibitem{rk2} All  prefactors that depend on the geometry of the sample and on the strength and shape of the photon source have been absorbed in the definitions of ${\cal D}^{(i,c)}$.
\end{thebibliography}
    \end{document}